\documentclass[aps,preprint,amssymb,12pt,floatfix]{revtex4}
\usepackage{subfigure}
\usepackage[pdftex]{graphicx}
\usepackage{rotating}
\usepackage{epstopdf}
\usepackage{color}    
\usepackage{amsmath,amsthm}
\begin{document}

\title{Refolding dynamics of stretched biopolymers upon force-quench}
\author{Changbong Hyeon$^{1*}$, Greg Morrison$^{2,3}$, David L. Pincus$^3$ and D. Thirumalai$^{3,4}$}
\thanks{To whom correspondence should be addressed. Email: thirum@umd.edu, hyeoncb@cau.ac.kr}
\affiliation{$^1$Department of Chemistry, Chung-Ang University, Seoul 156-756, Republic of Korea\\
$^2$School of Engineering and Applied Science, Harvard University, Cambridge, Massachusetts, 02138, USA\\
$^3$Biophysics Program, Institute For Physical Science and Technology, University of Maryland, College Park, Maryland 20742, USA\\
$^4$Department of Chemistry and Biochemsitry, University of Maryland, College Park, Maryland 20742, USA}
\date{\today}
\begin{abstract}
Single molecule force spectroscopy methods can be used to generate folding trajectories of biopolymers from arbitrary regions of the folding landscape.  We illustrate the complexity of the folding kinetics  and generic aspects of the collapse of RNA and proteins upon force quench, using simulations of an RNA hairpin and theory based on the de Gennes  model for homopolymer collapse.  The folding time, $\tau_F$, depends asymmetrically on $\delta f_S = f_S - f_m$ and $\delta f_Q = f_m - f_Q$ where $f_S$ ($f_Q$) is the stretch (quench) force, and  $f_m$ is the transition mid-force of the RNA hairpin.  
In accord with experiments, the relaxation kinetics of the molecular extension, $R(t)$, occurs in three stages: a rapid initial  decrease in the extension is followed by a plateau, and finally an abrupt reduction in $R(t)$ that occurs as  the native state is approached. 
 The duration of the plateau   increases as  $\lambda =\tau_Q/\tau_F$ decreases (where $\tau_Q$ is the time in which the force is reduced from $f_S$ to $f_Q$). Variations in the mechanisms of force quench relaxation as $\lambda$ is altered are reflected in the experimentally measurable time-dependent entropy, which is computed directly from the folding trajectories. 
An analytical solution of the de Gennes model under tension reproduces the multistage stage kinetics in $R(t)$.   The prediction that the initial stages of collapse should also be a generic feature of polymers is validated by simulation of the kinetics of toroid (globule) formation in semiflexible (flexible) homopolymers in poor solvents upon quenching the force from a fully stretched state. Our findings give a unified explanation for multiple disparate experimental observations of protein folding.
\end{abstract}

\maketitle

The folding of RNA molecules \cite{HyeonBC05} and proteins \cite{OnuchicCOSB04,Dill08ARB,Shakhnovich06ChemRev,Thirum95JPI} should be thought of as dynamic changes in the distribution of conformations that result in collapse, the formation of intermediates, and barrier crossings to reach the folded structures. 
Such a statistical mechanical perspective of the folding process is finding support in single molecule measurements that manipulate the structural ensembles using an external mechanical force ($f$) \cite{FernandezSCI04,Schlierf09AngChem,TinocoBJ06}.
The development of the force-clamp methods \cite{FernandezSCI04,TinocoBJ06}, which allow the application of a constant force to specific locations on a biomolecule,
have made it possible to explore in the most straightforward manner the folding of proteins and RNA initiated from regions of the folding landscape that are inaccessible in conventional ensemble experiments.   
The use of an initial stretching force, $f_S$, which fully unfolds the biomolecule, followed by a subsequent quench to a sufficiently low force, $f_Q$, which populates the Native Basin of Attraction (NBA), holds the promise of unearthing all aspects of the folding reaction, including the dynamics of the collapse process and the extent to which the folding pathways are heterogeneous. 

Although there is great heterogeneity in the folding trajectories, Fernandez and Li \cite{FernandezSCI04} noted that upon the quench $f_S\rightarrow f_Q$, ubiquitin (Ub) folded in three stages as reflected in the time-dependent changes in the extension, $R(t)$. Typically, after a very rapid decrease in $R(t)$, there is a long plateau in $R(t)$ that is suggestive of the formation of metastable states, which we term Force-Induced Metastable Intermediates (FIMI's).  Finally, $R(t)$ decreases sharply in the last stage as the extension corresponding to the native state is reached.  Although it is not emphasized, a similar behavior has been observed in the folding of TAR RNA (see Fig.3A in \cite{TinocoBJ06}) upon abrupt decrease in $f$, even though the forces used in Atomic Force Microscopy (AFM) experiments on Ub and Laser Optical Tweezer (LOT) experiments on the RNA differ greatly.  In contrast, it has been argued that Ub folds in a discrete manner based on refolding initiated by decreasing force slowly using AFM \cite{Schlierf09AngChem}, and not in the way described in \cite{FernandezSCI04}.  The seemingly contradictory conclusions on the refolding of an initially extended Ub require a full theoretical description of the molecular events that ensue when force is decreased from $f_S$.

In order to quantitatively describe the results of single molecule measurements, the interplay between a number of factors have to be taken into account.  The variables that can be controlled in experiments are $f_S$, $f_Q$, and $v_L$, the speed with which
the position of transducer is contracted (over a length $\Delta L$)  
 or equivalently, $r_c$, the rate of force contraction.  
Because the routes navigated
by the biomolecule will depend on the precise protocol used \cite{TinocoBJ06,Li07PNAS}, these factors must be taken into account when comparing different experiments, even if the biomolecule under consideration is the same.  For a system whose global folding obeys two-state kinetics, the two relevant time scales are the force relaxation time, $\tau_Q$ (with $\tau_{Q} \approx \Delta L/v_L$ 
or equivalently, $\tau_Q\approx (f_S-f_Q)/r_c$), and the overall folding time, $\tau_F(f_S,f_Q)$.
The dynamics of the folding trajectories, $R(t)$, should depend on  $\lambda = \tau_{Q}/\tau_F$. 
If $\lambda \ll 1$ then folding takes place far from equilibrium, whereas in the opposite limit refolding occurs under near-equilibrium conditions. Thus, by altering the force-quench protocol it is possible to control the collapse process and the folding routes for a given protein or RNA, even if $f_Q$ and $f_S$ are fixed.  


\begin{figure}[ht]
\includegraphics[width=6.00in]{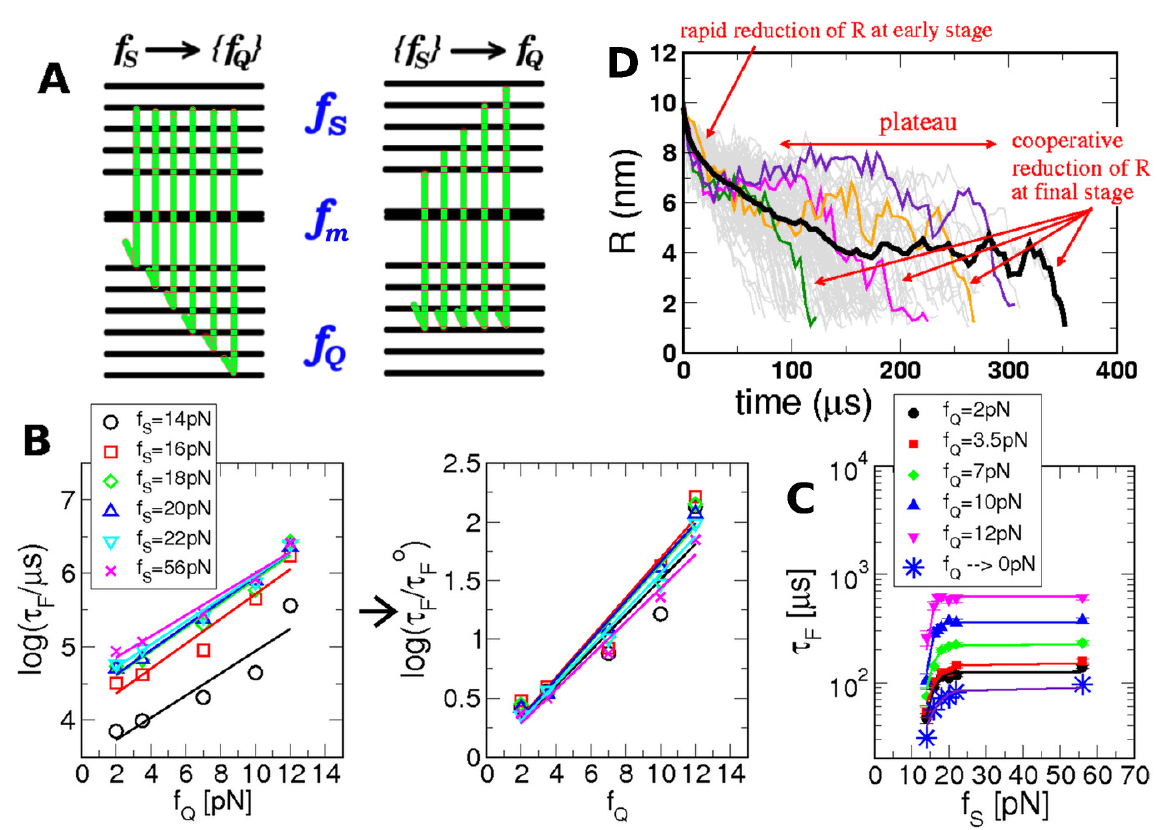}
\caption{Force-quench kinetics of P5GA hairpins. 
{\bf A}. 
Refolding from a fixed $f_S$ to varying $f_Q$ (left) or from varying $f_S$ to a fixed $f_Q$ (right). 
{\bf B}. $\tau_F$ versus $f_Q$ starting from different $f_S$.  
{\bf A}. Fit to the Bell equation gives $\Delta x_{U\rightarrow N}^{\ddagger}\approx (0.6-0.7)$ nm. The extrapolated value $\tau_F^o$ to zero quench force is plotted in {\bf C} ($\ast$ symbols). 
{\bf C}. $\tau_F$ versus $f_S$ for each value of $f_Q$. 
The variations of $\tau_F$ with $f_S$ are fit to Eq.\ref{eqn:tau_F}.
From the fit, 
we find ($\Delta x_{UN}/\mbox{nm}$,$\tau(f_Q)/\mu\mbox{s})=$(1.5, 89.7), (2.5, 126), (1.8, 150), (2.6, 224), (3.8, 357), (4.4, 622) for $f_Q=$0, 2, 3.5, 7, 10, 12 pN, respectively. 
{\bf D}. Time traces of the molecular extension, $R(t)$, upon force quench. 
Multiple time traces are plotted in grey, and the averaged time trace is shown in thick black. A few representative time traces are  in color. 
The ensemble of time traces are shown for $f_S=70$ pN and $f_Q=3.5$ pN. 
\label{Fig1.fig}}
\end{figure}

Here, we explore quantitatively the scenarios that  emerge for the folding of a RNA hairpin as a function of $f_Q$ and $f_S$. 
The value of $\lambda$ is controlled by altering $\tau_F(f_Q,f_S)$ in this work, rather than by adjusting the force contraction timescale $\tau_Q$, as in Refs \cite{FernandezSCI04,Schlierf09AngChem}.  We place 
particular emphasis on the general features of the physics underlying the early stages of length contraction. 
The  folding trajectories for a RNA hairpin exhibit a long plateau of varying length in $R(t)$, associated with the development of a FIMI. More generally, we show that the folding trajectories can be controlled by the choice of an appropriate force-quench protocol, which is succinctly expressed in terms of $\lambda$.  The origin of the long-lived FIMIs when $\lambda$ is small is explained by adapting de Gennes' picture of the kinetics of a coil to globule transition in flexible homopolymers.  The combination of simulations and theory shows that the formation of long-lived FIMIs should be a generic feature of any biomolecule that adopts a compact structure, which implies that the polymeric nature of RNA and proteins, along with interactions that induce a globular state for sufficiently low $f_Q$, will generically 
lead to a long plateau in $R(t)$ as long as $\lambda$ is small. These conclusions are supported by explicit simulations of force-quench studies of a flexible homopolymer in a poor solvent and a semiflexible chain model for DNA, in which the compact state adopts a toroidal structure. Our theory, that emphasizes $\lambda$ as the relevant variable in determining the folding routes, naturally explains the apparent differences in the interpretation of force-quench folding of Ub by different groups \cite{FernandezSCI04,Schlierf09AngChem}. \\

\noindent {\bf RESULTS}\\

{\bf Asymmetry in the force-quench kinetics : } 
Force-quench  refolding of a RNA hairpin (or a protein) can be initiated using two distinct modes.  In the first, 
the RNA is equilibrated at an initial $f_S>f_m$, where $f_m$ is the transition force at which the probabilities of being folded and unfolded are equal, and subsequently the force is quenched to various $f_Q$ values ($f_S\rightarrow \{f_Q\}$), with $f_Q<f_m$ (Fig.1).  We use simulations of the P5GA hairpin  \cite{Hyeon08PNAS}, with $f_m=14.7$ $pN$ at $T=300$ K (see Methods) to extract the dependence of $\tau_F$ on $f_Q$ from the dynamics of folding trajectories (Fig.1D and Fig.2). 
Similarly, refolding can also be initiated by varying $f_S(>f_m)$ and reducing the force to a single $f_Q (<f_m)$ ($\{f_S\}\rightarrow f_Q$) (see Fig.1A).  

The mean refolding time ($\tau_F=\frac{1}{N}\sum_{i=1}^N\tau_F(i)$) for different values of $f_Q$ is fit to a Bell equation \cite{BellSCI78,FernandezSCI04,Li06PNAS} (Fig.1B), 
  $\tau_F(f_Q,f_S)=\tau_F^o(f_S)e^{f_Q\Delta x_{U\rightarrow N}^{\ddagger}/k_BT}$
where $\tau_F^o(f_S)$ is the folding time at $f_Q=0$, and $\Delta x_{U\rightarrow F}^{\ddagger}$ is the distance between the transition state and the free energy minimum in the unfolded basin of attraction (UBA), which assumes that $R$ is a good reaction coordinate.  
The slight upward curvature in $\log(\tau_F)$ as a function of $f_Q$ is due to the dependence of $\Delta x_{U\rightarrow N}^{\ddagger}$ on $f_Q$ \cite{HyeonBJ06,Hyeon07JP,Dudko06PRL}.
When $\tau_F$ is rescaled by $\tau_F^o(f_S)$, the refolding times at different  $f_S$ values nearly collapse onto a single curve 
(see the panel on right hand side in the Fig.1B). 
 It follows that the time scale for crossing an effective free energy barrier (in the late stages of folding) is solely determined by $f_Q$, and is independent of $f_S$. 
 The microscopic model in \cite{Dudko06PRL,Dudko08PNAS} provides an excellent fit to the simulation data (see also Fig.S2 and the caption therein). Here, we use the Bell equation to provide a physically simpler picture of the effect of force on the free energy profile.

Unlike the behavior of $\tau_F$ with varying $f_Q$, the dependence of $\tau_F$ on $f_S$ cannot be  explained using the Bell model.
We find that
$\tau_F$ decreases sharply with $f_S$ for $f_S<20$ pN, but
saturates to a limiting value (which is also dependent upon $f_Q$) if $f_S>20$ pN (Fig.1C).  
The saturation of $\tau_F$ for large $f_S$ suggests that the initial conformations of P5GA belong to the same structural ensemble (see the variations of the end-to-end distribution, $P_S(R)$, at each $f_S$ in Fig.S1A). 
At low $f_S$ ($f_m\lesssim f_S<20$ pN) the molecules are partitioned into different ensembles. 
The initial population of molecules in the UBA, determined by $f_S$, affects the average folding time. 
Since P5GA is two-state folder under force, the fraction of molecules in the UBA at $f_S$ is $\varphi_{UBA}(f_S)=1/[1+e^{(\Delta F^o_{UN}-f_S\Delta x_{UN})/k_BT}]$, where $\Delta F^o_{UN}=F_U^o-F_N^o$ and $\Delta x_{UN}=x_U-x_N$ are the difference in free energy and molecular extension between unfolded and folded conformations, respectively. 
By combining the $f_Q$-dependence of $\tau_F$ from Bell's model with $\varphi_{UBA}(f_S),$ we obtain a unified expression that describes force-quench folding time in both modes (see SI for details),
 \begin{equation}
	  \tau_F(\delta f_S,\delta f_Q)=\frac{\tau(0)}{1+e^{-\delta f_S\Delta x_{UN}/k_BT}}e^{-\delta f_Q\Delta x^{\ddagger}_{U\rightarrow N}/k_BT}
	  \label{eqn:tau_F}
	\end{equation}
where $\delta f_S=f_S-f_m$, $\delta f_Q=f_m-f_Q$, $\tau(0)=\tau_oe^{\Delta F^{\ddagger}_{U\rightarrow N}/k_BT}$ with $\tau_o$ being the time scale in the absence of activation barrier, and we have used $\Delta F^o_{UN}-f_m\Delta x_{UN}=0$. 
The fits using the two parameters 
$\tau_F(f_Q,f_S)(\equiv \tau(0)e^{-\delta f_Q\Delta x_{U\rightarrow N}^{\ddagger}/k_BT})$ 
and $\Delta x_{UN}$, provides an excellent description of the simulation data (Fig.1C). 
The analytic result in Eq.\ref{eqn:tau_F} also shows the inherent asymmetry in the two modes for initiating folding by force-quench. 
\\

\begin{figure}[ht]
\includegraphics[width=6.50in]{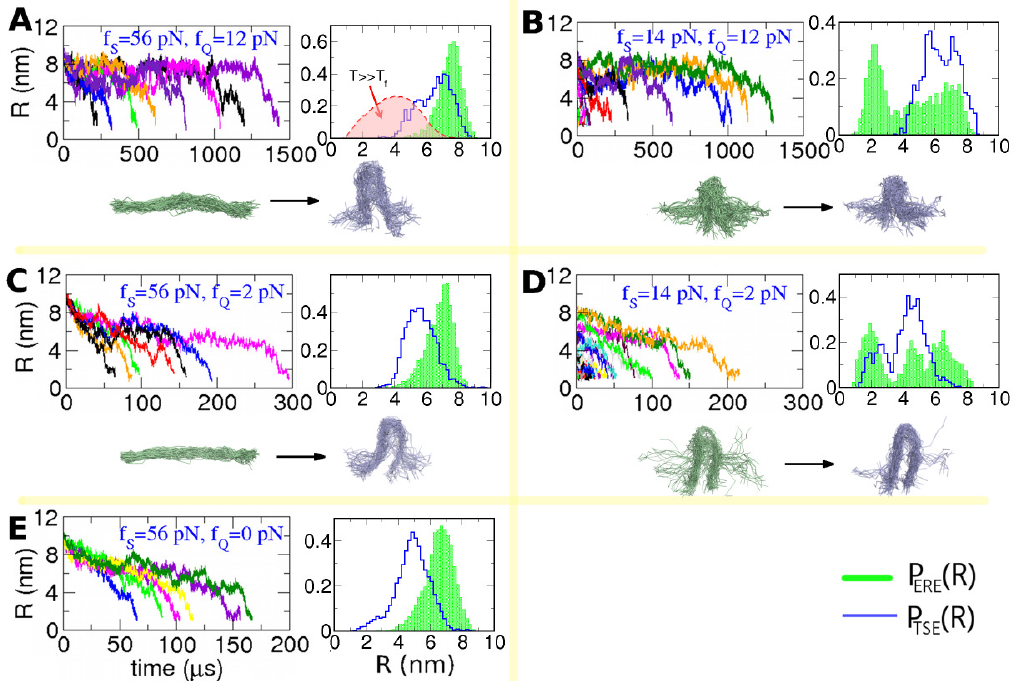}
\caption{Analysis of the folding trajectories associated with force-quench for various values of ($f_S, f_Q$). 
  {\bf A} : $f_S=56$ pN $\rightarrow f_Q=12$ pN, {\bf B} : $f_S=14$ pN $\rightarrow f_Q=12$ pN, {\bf C} : $f_S=56$ pN $\rightarrow f_Q=2$ pN, {\bf D} : $f_S=14$ pN $\rightarrow f_Q=2$ pN. {\bf E} : $f_S=56$ pN $\rightarrow f_Q=0$ pN.  $P_{ERE}(R)$ and $P_{TSE}(R)$ are shown in the right-hand side panels of {\bf A}-{\bf E}, and the corresponding structures are shown for the ERE (left) and TSE (right).
Thermally denatured ensemble $P_{TDE}(R)$ for $T\gg T_f$ 
is shown in {\bf A}. 
\label{Fig2.fig}}
\end{figure}

{\bf Non-equilibrium force-quench folding is reflected in $R(t)$:}  
The dramatic variations in the refolding time $\tau_F(f_Q,f_S)$  (Figs.1B and 1C) are reflected in the folding trajectories, $R(t)$, as $f_S$ and $f_Q$ are varied. 
In Fig.1D (see also Fig.2), which shows the ensemble of force-quench trajectories with $f_S= 70$ pN and $f_Q=3.5$ pN, we observe 
a rapid decrease in $R(t)$ in the early stages of folding (stage 1), followed by a slowly decaying plateau of varying length (stage 2). 
The second stage is followed by an abrupt reduction to 
$R\lesssim 2$ nm, corresponding to the hairpin formation \cite{Hyeon08JACS} \cite{FernandezSCI04,HyeonPNAS05,HyeonBJ06} (stage 3).  
The three stages of folding are well separated temporally, and the plateau length increases with increasing $f_Q$. 
Although multistage folding has been discussed in the context of T-jump protein folding \cite{Sali94Nature,CamachoPNAS93,Thirum95JPI}, the dramatic manifestation of timescale separation at an ``early" stage of protein folding is unique to force quench experiments. 
The nature of FIMI is different from that of a thermally collapsed intermediate state in that the FIMI state contains no intramolecular contacts (see below). 
The rapid increase in the folding times (Figs.1B and 1D) is directly linked to the long-lived FIMIs. 
The plateau persists until the system reaches the transition state, $R\approx R_{TS}=4$nm, where $R_{TS}$ is the position of transition barrier at $f=f_m$ (Fig.S3). 

The process by which RNA navigates the folding landscape in stage 2 of the collapse process, during which $R(t)$ is nearly a constant (Fig.1D), is assessed by monitoring the conformational changes in the hairpin. 
In order to illustrate these structural changes, we computed the distributions,  
$P_{ERE}(R)$ (ERE stands for entropically relaxed ensemble) ``immediately after'' the force-quench (between stages 1 and 2), and $P_{TSE}(R)$ (TSE denotes the transition state ensemble) that gives the distribution function ``immediately before'' the final decrease of $R(t)$ (between stages 2 and 3; see the SI for further discussion).  
The differences between $P_{ERE}(R)$ and $P_{TSE}(R)$ are illustrated in Fig.2.  As long as $f_S\gg f_m$, 
$P_{ERE}(R)$ consists of a set of homogeneously stretched structures, whereas the structural ensemble of hairpin loops with disordered ends become dominant for $f_S\lesssim f_m$ (see Fig.2A-2D).
For $f_S=56$ pN, $P_{TSE}(R)$ is peaked at smaller values of $R\approx (6-7)$ nm, and becomes broader with decreasing $f_Q$ (see Fig.2A, 2C, and 2E). 
The ensemble of structures that are sampled in this second stage are the FIMIs, which are analogues of minimum energy compact structures, postulated by Camacho and Thirumalai \cite{Camacho93PRL} for folding initiated by temperature quench. 
Comparison of $P(R)$ in Figs 2A, 2C and 2E with $P_{TDE}(R)$ (TDE denotes the thermally denatured ensemble; see the pink distribution in the Fig. 2A), show that the conformations in the FIMI are unlikely to be sampled during a typical refolding initiated by temperature quench \cite{Hyeon08JACS}.

At $f_S=14$ pN $\lesssim f_m$ (Fig.2B and 2D), 
$P_{ERE}(R)$ differs substantially from the distributions for $f_S=56$ pN $\gg f_m$ (Fig.2A, 2C, and 2E). 
The ERE has an approximately bimodal distribution, similar to the initial distribution $P_S(R)$ at $f=f_m$ \cite{HyeonBJ07,Hyeon08PNAS} (compare $P_{ERE}(R)$ in Fig.2B, 2D with Fig.S1 at $f_S=14$ pN), and is far more heterogeneous than the ensemble at $f_S=56$ pN. 
With $f_S\approx f_m$, the folding is initiated predominantly from molecules with a partially formed hairpin loop. 
In this case, the molecule refolds directly to the native state without a well-defined plateau in $R(t)$. 
A plateau in $R(t)$ will be manifested most clearly only when the force is quenched from a large $f_S(\gg f_m)$ and $f_Q(\lesssim f_m)$.  
The simulations also suggest that the lifetimes of the FIMIs depend on the protocol used to quench the force, which is determined by $\lambda$ (see Table I). 

The evolution from the ERE to the TSE, which occurs over a time scale determined by the length of the plateau, is associated with a search process leading to a nucleation step in the hairpin formation. 
We have shown previously, in a force quench simulations with $f_Q=0$, that the dynamics of the plateau is associated with a search for the native dihedral angles (trans$\rightarrow $ gauche transition) in the tetra-loop region 
\cite{HyeonBJ06,Hyeon08JACS}. 
However, the plateau formation in $R(t)$ is a generic feature of force-quench dynamics, and has also been observed in recent experiments on poly-ubiquitin, I27, and the PEVK protein from titin \cite{FernandezSCI04,Walther07PNAS}.  
Interestingly, the formation of long-lived FIMIs also occurs in force-quench dynamics of semiflexible chains and flexible homopolymers (see Discussion). 
\\

\begin{figure}[ht]
\includegraphics[width=6.00in]{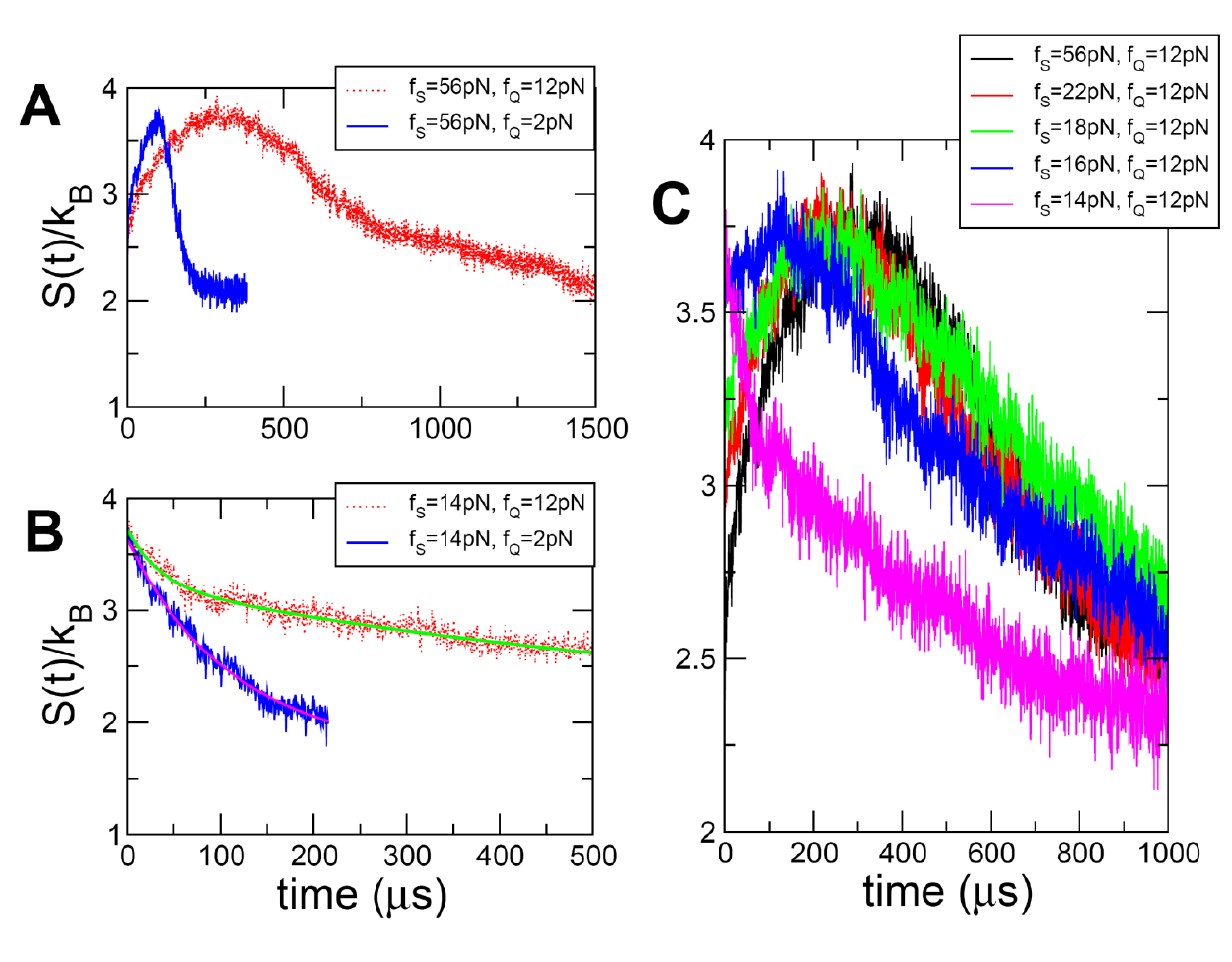}
\caption{The time evolution of the entropy, measured using Eq. \ref{TimeEntropyEq}.
{\bf A}. $S(t)/k_B$ from $f_S=56$ pN to $f_Q=12$ pN (red) and 2 pN (blue).
{\bf B}. $S(t)/k_B$ from $f_S=14$ pN to $f_Q=12$ pN (red) and 2 pN (blue).
{\bf C}. The time evolution of $S(t)/k_B$ for varying $f_S$ with fixed $f_Q=12$pN. In this panel, we take the running-average of $S(t)/k_B$ every 4.7 $\mu $s.
\label{Fig3PNAS.fig}}
\end{figure}

{\bf Link between entropy production and the collapse dynamics : } 
The transition from the stretched state to the folded state that occurs through FIMIs can be described using time dependent changes in the entropy. 
From the ensemble of force-quench time traces, $R(t)$, we  estimate the  
entropy of the molecule as a function of time by using 
\begin{equation} 
S(t)/k_B=-\int_0^{\infty} dR\, P(R,t)\log{P(R,t)},  \label{TimeEntropyEq}
\end{equation} 
where $P(R,t)$ is the end-to-end distribution function obtained directly from the ensemble of trajectories.  
At long times $P(R,t\rightarrow\infty)$ will reduce to the distribution for a given system. The time-dependent entropy is computed directly from the data generated from experiments or simulations, giving this approach a practical use, and thus captures the non-equilibrium behavior in all force regimes and over all time scales. Note that Eq. \ref{TimeEntropyEq} makes no assumptions about a particular model for $P(R,t)$ but is strictly valid only if $R$ can adequately describe the conformation of the molecule of interest. 
Fig.3 shows $S(t)$ computed for four different pairs of $(f_S, f_Q)$.  
Interestingly, changes in $S(t)$ are nonmonotonic when the system begins from a high stretch force ($f_S=56$ pN, Fig.3A),
whereas the $S(t)$ decreases monotonically upon force-quench from $f_S=14$ pN (Fig.3B).  
The quantitative analysis using $S(t)$ clarifies the dynamics in the ensemble of time traces (Fig.2).
The evolution of the end-to-end distribution, $P_S(R)\rightarrow P_{ERE}(R)\rightarrow P_{TSE}(R)\rightarrow P_{NBA}(R)$, is due to changes in the dynamic evolution of the entropy (Eq. \ref{TimeEntropyEq}) which progresses from a low (stretched) $\rightarrow$medium (ERE) $\rightarrow$high (FIMIs, TSE) $\rightarrow$low (NBA) value as the hairpin folds, especially when the initial ensemble of states is fully stretched ($R\gg 4$nm in Fig.S2).  
At $f_S=14$ pN, $P_S(R)$ is bimodal and $S(t)$ with a maximum value  at $t=0$, and decreases  monotonically  as   
$S(t)/k_B\sim \Phi e^{-t/\tau_f}+(1-\Phi)e^{-t/\tau_s} \label{DoubleExpEq}$
where $\Phi$, $\tau_f$ and $\tau_s$ are $f_Q$-dependent for the same value of $f_S$. The time constants $\tau_f$ and $\tau_s$ correspond to collapse of  fast- and slow-track molecules in the multidimensional free energy landscape (with $\tau_f<\tau_s$), and $\Phi$ is the fraction of fast collapsing molecules \cite{HyeonBC05}.   The fast and slow tracks are associated with the dynamics of collapse, starting from within the NBA ($R<4$nm) and UBA ($R>4$nm), respectively.   
We find $\Phi =0.29$, $\tau_f=33.8$ $\mu s$, $\tau_s=715$ $\mu s$ for $f_Q=12$ pN, and $\Phi =0.47$, $\tau_f=81.3$ $\mu s$, $\tau_s=181$ $\mu s$ for $f_Q=2$ pN.
A smaller value of $f_Q$ increases the fraction of conformations in the folding route starting from NBA to the native state.  
The force-quench dynamics from the force-denatured ensemble (FDE) with $f_S=14$ pN is an example of the kinetic partitioning mechanism \cite{HyeonBC05}, and shows that the mechanisms of hairpin formation can be surprisingly complex \cite{TinocoBJ06}.

The transition from the monotonic to non-monotonic decay of $S(t)$ occurs at $f_S\approx 16$ pN (Fig.3C), and the time scales for $S(t)$ to attain its maximum value are similar for $f_S=18-56$ pN. However, the entropies of the initial structural ensemble, $S(0)$, are different and decrease as $f_S$ increases.  
This finding provides insights into why $\tau_F$ saturates for $f_S\geq 18$ pN in Fig.1C, since the P5GA hairpin is predominantly in the UBA for $f_S>f_m+\delta f_{1/2}=14.7+1.2$ pN $\approx 16$ pN, where $\delta f_{1/2}$ is the half-width of force fluctuations around $f_m$ (see SI for details). 
 \\ 

\begin{figure}[ht]
\includegraphics[width=5.00in]{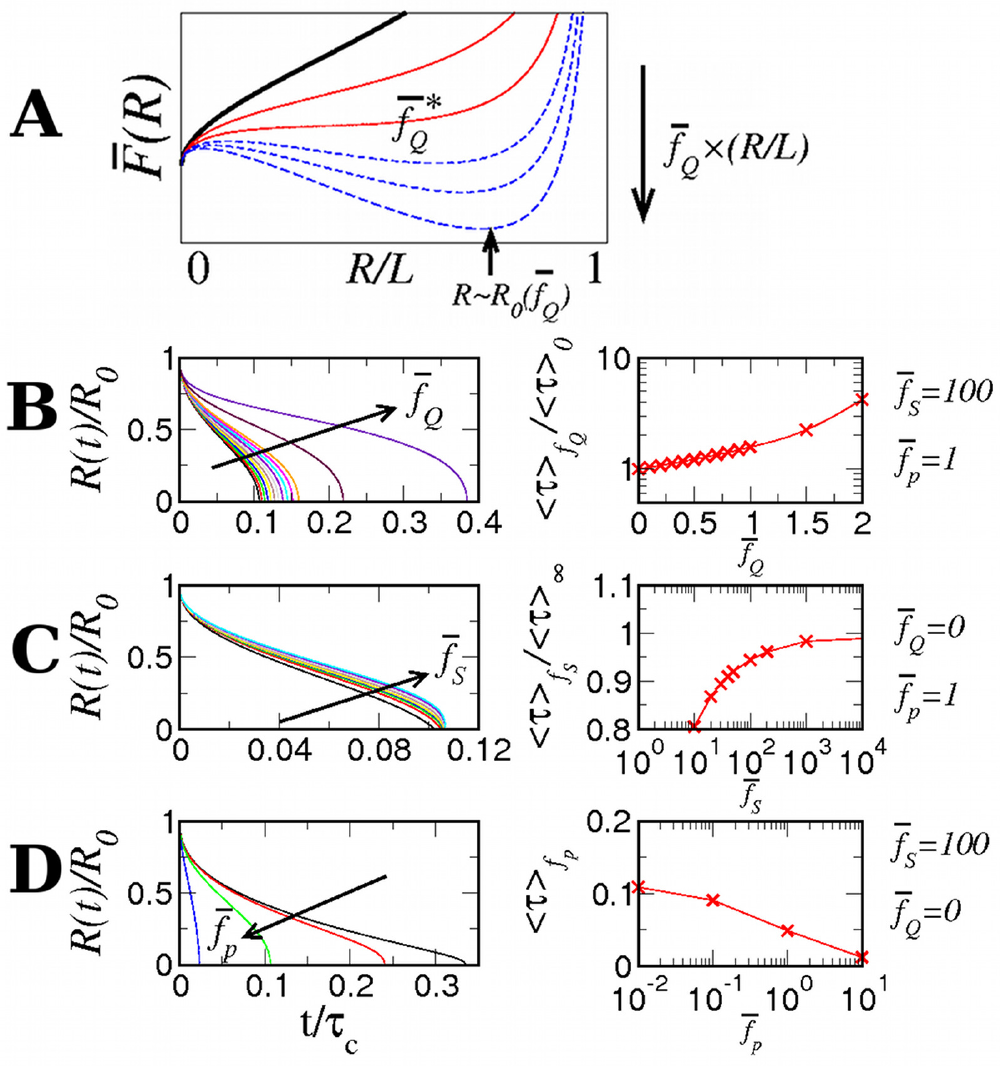}
\caption{The expanding sausage model for force quench dynamics. 
{\bf A}. The free energy profile as a function of $R$ for increasing tension. 
The free energy profile $F(R)$ is plotted in dimensionless form using $\overline{F}(R/L)=\frac{F\tau_c}{\eta L^3}=2\left(\frac{R}{L}\right)^{1/2}+\overline{f}_p\frac{R^2}{4L^2}\left(\frac{3-2R/L}{1-R/L}\right)-\overline{f}_Q\left(\frac{R}{L}\right)$.   
Two stable free energy minima occur at $R=0$ and $R=R_0(f_Q)\lesssim L$ when $\overline{f}_Q>\overline{f}_Q^{\ *}$. 
The profiles show that $R^{TS}$ decreases and $R_0$ increases with increasing $f$. 
In {\bf B}, {\bf C}, and {\bf D} the dynamics of $R(t)$ (left) and the average refolding time $\langle\tau\rangle=\tau_F$ (right) are shown for varying $\overline{f}_Q$, $\overline{f}_S$, and chain flexibility ($\overline{f}_p=k_B T/l_p$), respectively.  
\label{Fig4PNAS.fig}}
\end{figure}

{\bf The expanding sausage model under tension captures the physics of force-quench refolding of biopolymers  : } 
Although the non-equilibrium behavior of the hairpin, reflected in $R(t)$ and the time scales for hairpin formation is shown for P5GA (Fig.1 and 2), the dynamics of contraction of the molecule upon rapid quench should be a generic phenomenon, provided $\lambda$ is small and a driving force leads to the development of a compact structure for sufficiently low $f_Q$. In order to illustrate the generality of the FIMI that gives rises to the second stage of contraction (Fig.1, 2 and ref. \cite{FernandezSCI04}), we provide a theoretical description by adapting de Gennes' model \cite{deGennes85JPL} for collapse kinetics of a flexible coil. 
Using an ``expanding sausage model,'' 
de Gennes envisioned that the collapse (upon temperature quench to $T=T_{\Theta}-\Delta T< T_{\Theta}$, the Flory $\Theta$-temperature) progresses from a randomly distributed string of thermal blobs of size $\xi (=ag^{1/2}$, where $g[=(T_{\Theta}/\Delta T)^2]$ is the number of monomer constituting each blob), to a sausage-like collapsed structure. 
The driving force in this picture (see Fig.S4A) is the minimization of the surface tension in a poor solvent ($T<T_{\Theta}$). 

To describe the second stage in the force quench dynamics (or more generally for any biopolymer that undergoes a stretch $\rightarrow$ collapse transition) using the generalization of expanding sausage model, 
we write the free energy under tension as 
\begin{align}
{F(R)}{}=
2\frac{k_B T}{\xi^{2}}\sqrt{\pi\Omega}R^{1/2}- f_QR
+k_B T\frac{L}{l_p}\int_0^{R/L}dx\left[\frac{1}{4(1-x)^2}-\frac{1}{4}+x\right].
\label{eqn:WLC-F}
\end{align}
where the first term is the driving force for collapse ($F_C=\gamma A$ where A is the surface area of a thermal blob). 
We rewrite $F_C\sim 2k_BT(\pi \Omega R)^{1/2}/\xi^2$, where the surface tension for a thermal blob is $\gamma  \approx k_BT/\xi^2$, and $\Omega$ is the volume of the sausage, assumed to be constant throughout the collapse process \cite{deGennes85JPL}. 
The second term in Eq. \ref{eqn:WLC-F}, $F_f=-f_QR$, accounts for the mechanical work. 
The third term in Eq. \ref{eqn:WLC-F}, which enforces chain inextensibility under the applied tension, is required to incorporate the sharp increase of mechanical force at $R\approx L$ observed in P5GA hairpin (see Fig S1B).  A simplified version of the sausage model, where inextensibility is not taken into account, is presented in the SI.  
The details of the free energy (Eq.\ref{eqn:WLC-F}) are not expected to change the qualitative picture of the collapse kinetics that emerges using the expanding sausage model to describe force-quench dynamics, as long as each term is taken into account.  
The effect of the stretch force $f_S$ on the folding kinetics is introduced through the initial equilibrium extension $R_0$, which for $f_S\gg k_BT/l_p(\equiv f_p)$ is 
$R_0(f_S)\approx L\left(1- \sqrt{f_p/4f_S}\right)\lesssim L$. 

We equate the rate of free energy reduction of the system, 
$\dot{F}(R)$, with the entropy production due to dissipation,
$T\dot{S}(R)\approx \eta\times R\times \dot{R}^2$ \cite{deGennes85JPL},
where we have used the Stokes-Einstein relation and a damping force $f\approx -\zeta v\approx -\eta R v$. The friction coefficient $\zeta\approx \eta R$ where $R$ is the size of collapsed polymer in the longitudinal direction, and $v(\approx \dot{R})$ is the velocity.
By equating $\dot{F}(R)\approx-T\dot{S}(R)$ we get 
\begin{align}
t/\tau_c&=\int^{R_0/L}_{R/L}\frac{x\ dx}{x^{-1/2}-\overline{f}_Q+\overline{f}_p\left(\frac{1}{4(1-x)^2}-\frac{1}{4}+x\right)}
\label{eqn:tf_WLC}
\end{align}
where $R_0/L\approx 1-\sqrt{f_p/4f_S}$ for ${f}_S\gg {f}_p$ the forces are scaled as $\overline{f}_{\alpha}=f_{\alpha}\tau_c/\eta L^2$ for $\alpha=S$, $Q$, and $p$, and with $\tau_c\equiv \frac{\eta\xi^2L^{5/2}}{k_BT\sqrt{\pi\Omega}}$. 
The ratio $R(t)/R_0\equiv \Sigma(t)$, which is interpreted as the survival probability of the unfolded state upon force quench $f_Q$, can be determined by a numerically integrating Eq. \ref{eqn:tf_WLC} (see the left column of Figs.4B-D).
The average folding time (or mean first passage time) is 
$\tau_F(\{f_{\alpha}\})=\langle\tau\rangle_{\{f_{\alpha}\}}=\int^{\tau^*}_0dt\  \Sigma(t;\{f_{\alpha}\})$,
where $\tau^*$ satisfies $R(\tau^*)=0$. 
The results of $\tau_F$ for various combination of ($\overline{f}_Q$, $\overline{f}_S$, $\overline{f}_p$) are given 
in the right column of Fig. 4B-D. 
In agreement with our simulation results (see also \cite{HyeonPNAS05,HyeonBJ06,Hyeon08JACS}), as well as force quench experiments on poly-Ub \cite{FernandezSCI04}, the sausage model for WLC shows a sharp decrease in 
$R(t)$ at an early stage of the force-quench dynamics, after which a plateau in $R$ of variable length develops, before an abrupt collapse to $R=0$ (see Fig.2). 
A long time plateau is observed for a stiffer biopolymer (small $f_p$), and for larger $f_Q$ (Figs.4B and 4D).
In agreement with the more detailed P5GA simulations, 
$\tau_F$ increases exponentially at small quench forces ($\overline{f}_Q\lesssim 1$), with a slight upward curvature for larger $f_Q$ (compare Fig.4B). 
The dependence of $\tau_F$ on the stretch force $f_S$ is typically small, and saturates when $\overline{f}_S\gtrsim 10^3$ (compare Figs.1C and 4C). 

A word of caution is appropriate. 
Eq.\ref{eqn:tf_WLC}, which is derived in the absence of fluctuating force (noise term in the Langevin equation), is valid only for $\overline{f}_Q<\overline{f}_Q^{\ *}$ where $\overline{f}_Q^{\ *}$ is the force beyond which a finite free energy barrier for refolding develops (the dotted energy profiles in Fig.4A). 
Above $\overline{f}_Q=\overline{f}_Q^{\ *}$, there is a barrier to refolding, $\delta \overline{F}^{\ddagger}$, and a valid solution to Eq. \ref{eqn:tf_WLC} cannot be found for all $t>0$. 
De Gennes' approach \cite{deGennes85JPL}
 is not applicable for rate processes \cite{Chandrasekhar43RMP} that involve crossing a free energy barrier. 
Physically, the residence time in the FIMI state becomes infinite in the absence of noise. 
The analog of $\overline{f}_Q^{\ *}$ was referred to as ``spinodal'' or ``coercive field'' in Ref. \cite{Binder73PRB} in which the nonequilibrium relaxation of spin systems was studied. 
Therefore, to account for the force quench dynamics when $\overline{f}_Q>\overline{f}_Q^{\ *}$, we use the Kramers theory for barrier crossing, assuming that $R$ (under tension) is a good reaction coordinate, 
an approach that is consistent with other estimations of the transition time \cite{Yoshinaga08PRE,Halperin00PRE}. In so doing, it is assumed that under tension ($f_Q\neq0$) that results in the suppression of internal fluctuations $R$ is a good reaction coordinate. 
The refolding time $\tau_F$ scales with the effective height of the free energy barrier as $\tau_F\sim \exp{(\delta F^{\ddagger})}$ where $\delta\overline{F}^{\ddagger}=\overline{F}(R^{TS}/L)-\overline{F}(R_0/L)$. 
For $\overline{f}_Q\gg\overline{f}_Q^{\ *}$ one can show (see SI for details) 
\begin{equation}
\tau_F\sim \exp\left[\left(\frac{1-2\overline{f}_p^2}{\overline{f}_Q}-2+\overline{f}_Q\right)+\left(2-\overline{f}_Q+\frac{3}{2}\overline{f}_p\right)\sqrt{\frac{\overline{f}_p}{\overline{f}_Q}}\ \right]. 
\label{eqn:barrier_tau}
\end{equation}
For small $\overline{f}_Q$ or increasing flexibility ($\overline{f}_p\gg 1$), $\tau_F$ in Eq.\ref{eqn:barrier_tau} deviates significantly from Bell's expression [$\tau_F\sim \exp{(\overline{f}_Q)}$]  \cite{HyeonBJ06,Hyeon07JP,Dudko06PRL}.  We stress that Eq.\ref{eqn:barrier_tau} is only approximate and is shown to emphasize that activated transitions (in contrast to Eq. 4) dominates the approach to the ordered state. 

While additional factors such as topological constraints due to self-avoidance \cite{Grosberg88JP} or other interactions may alter the dynamic characteristics (for instance the dependence of $\tau_F$ on $f_S$ or $f_Q$) the generic feature of the emergence of a FIMI state upon force quench, also evidenced in the force-quench collapse dynamics of polymer chain under poor solvent conditions (see Discussion),  is satisfactorily captured by the simple model.
\\
\begin{figure}[ht]
\includegraphics[width=6.00in]{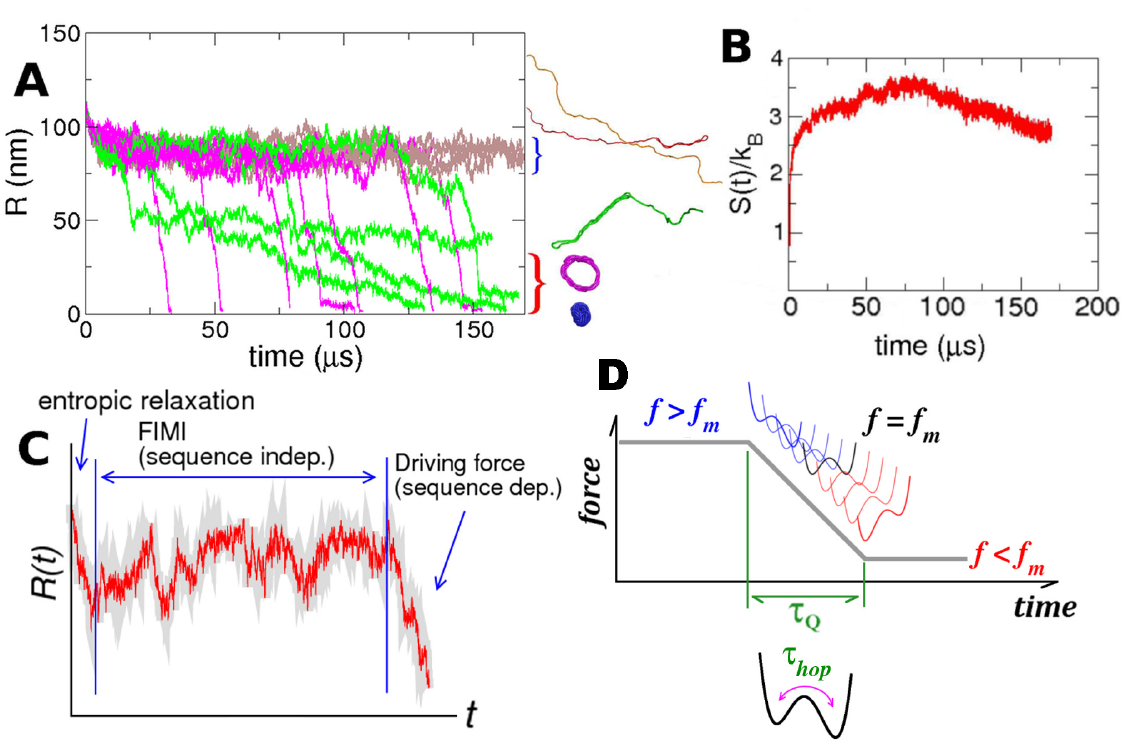}
\caption{Polymer collapse in a poor solvent upon force quench. 
{\bf A}.  The collapse of a WLC in a poor solvent.  
{\bf B}. Toroidal, single racquet, or multiple racquet structures are formed in various trajectories. 
{\bf C}. The entropy production  using Eq. \ref{TimeEntropyEq}, displaying the low$\to$high$\to$low pathway from high $f_S$ to low $f_Q$. 
{\bf D}. Linear reduction in force over a period, $\tau_Q$, during which the free energy gradually changes from a profile with $f>f_m$ to one with $f<f_m$. 
Depending on the time scale for hopping between the native and unfolded basins of attraction, $\tau_{hop}$, in comparison to $\tau_Q$, the pattern of relaxation dynamics upon force quench can be greatly affected.   
\label{Fig5PNAS.fig}}
\end{figure}

\noindent {\bf DISCUSSION}\\

{\bf{Implications for experiments:}}  
Our finding that the very nature
of folding can be altered by the depth of quench as well as the rate
of quench readily explains the different conclusions reached in the
refolding of Ub \cite{Schlierf09AngChem}.  As shown in Figs.1 and 2, the lifetime of the
FIMIs, leading to the plateau in $R(t)$, arises only if
the ratio $\lambda =\tau_{Q}/\tau_F(f_Q,f_S)$ is small. Under the
experimental conditions in ref \cite{FernandezSCI04}, $\lambda \approx 0.15$, 
whereas in the more recent AFM experiments, $\lambda \approx 32$ \cite{Schlierf09AngChem} (see Table I for details).  Consequently, the experiments by Schlierf and Rief \cite{Schlierf09AngChem} are
closer to equilibrium conditions than the force quench refolding
reported in \cite{FernandezSCI04}.  
We predict that the apparent differences between the conclusions reached in \cite{FernandezSCI04} and \cite{Schlierf09AngChem} are solely due to differing values of $\lambda$. 
The generality of these findings can
be further appreciated from the study of Li \emph{et. al.} on TAR RNA ($\lambda=0.1$, see Table I) that showed
 evidence for the development of a plateau (see Fig.3A in ref. \cite{TinocoBJ06}), although its
importance was not discussed.
Our findings also imply that only by varying $\lambda$ over a wide range it is possible to discern the multiple stages in folding, and firmly establish a link between collapse and folding transitions.\\

{\bf FIMIs are a generic feature of force-quench dynamics :} 
The sharp transitions in $R(t)$,
from its value in the second stage to the native compact structure in the third stage is reminiscent of a weak first order phase transition. 
Analysis of $P_{TSE}(R)$ for P5GA (Fig.2A) show that the formation of the hairpin loop from a FIMI is the key nucleation event in bringing the stretched RNA into its native-like hairpin structure. Because the simulations on RNA hairpins and the expanding sausage model suggest the long plateau only depends on the characteristic time scales associated with the folding polymer and the experimentally controlled variables in force quench, a similar behavior should be observed even in polymers that do not have a unique native structure, but can undergo a transition to a globular state.
To this end, we consider the dynamics of $R(t)$ in a homopolymers (flexible and worm-like chains) in a poor solvent upon force quench.  
The simulation details are in the SI.  


{\it Collapse of Semiflexible chain : } 
In the absence of force, the thermodynamically stable state of a semiflexible chain in poor solvent is a toroid (see Fig.5A). 
We consider the collapse of a 
stiff wormlike chain (bending rigidity $k_b=80 k_B T$, see Eq.(18) in SI) 
that is initially equilibrated at $f=f_S=83$ pN, and the force is suddenly quenched to $f_Q=4$ pN.  
From the time traces of $R(t)$ in Fig.5A, we surmise that the transition to the toroidal structure occurs in three stages, just as for the P5GA hairpin and Ub \cite{FernandezSCI04}.  
Within our simulation times ($\lesssim 170$ $\mu $s), some molecules adopt toroidal (magenta) or racquet structures (green), while others (brown) fluctuate around large values of $R$ without adopting any ordered structure.  
These findings for the formation of toroids for the WLC are in quantitative accord with 
experiments on the dynamics of force quench collapse of $\lambda-$DNA, which showed that toroid formation occurs in a stepwise, quantized manner \cite{Fu06JACS}, and is in agreement with the forced-unfolding of lattice model of proteins \cite{Klimov99PNAS}. 

The entropy production, $S(t)$, of a collapsing WLC is shown in Fig.5B 
only for those molecules that form an ordered structure (the trajectories grouped in red curly bracket in Fig.5A (or see Fig.S5 for complete set of trajectories)). 
The low entropy of the initial, highly stretched state increases at first, until it attains a maximum value, corresponding to an ensemble dominated by the FIMIs, after which the entropy slowly decays to a value in the more ordered, low entropy state.  This finding is in agreement with the three-stage collapse dynamics of the simpler RNA hairpin with a well-defined folded structure.

{\it Collapse of flexible polymers in a poor solvent :} 
To further demonstrate that bending rigidity in the semiflexible chain plays no special role in leading to FIMIs we have considered the stretch$\rightarrow$globule transition of a flexible hompolymer in a poor solvent under force quench conditions (see SI text for details). 
The final globular structure, in the absence of force, forms continuously from the ends of the chain (a pearl-necklace ensemble), with no evidence of a plateau in the collapse (see Fig.S6A).  However, the continuous collapse transition becomes cooperative and weakly first order by choosing a large quench force $f_Q=75$ pN (Fig.S6B), due to the formation of FIMIs. 

These two examples for homopolymers and the results for RNA hairpin show that the value of $\lambda$, and not the specific architecture of the folded structure, dictate the collapse in the intermediate stages. 
The expected changes in $R(t)$ of a biomolecule (or a homopolymer in a poor solvent) upon a rapid $f_S\rightarrow f_Q$ quench (small $\lambda$) must occur in multiple stages (Fig.S5) with stages 1 and 2 being determined by entropy growth (stage 1).  The search for the folding nuclei \cite{Guo97FD} that can further drives structure formation (stage 3) and sequence-dependent effects arise in the final stages of folding (Fig.5C). 

If the force is quenched rapidly, the biomolecule will be far from equilibrium when the first nucleation event occurs, which will drastically change the dynamics of the folding process.  Folding occurs in a near equilibrium manner either if the force is relaxed slowly (Fig. 5D), so that the internal modes of the biomolecule equilibrate or if  the quench force is sufficiently large that there is a significant barrier to folding. 
For a slow decrease in the force, the relevant timescale are $\tau_Q=\Delta L/v_L$, whereas the relevant timescale for the barrier crossing event will be $\tau_{hop}(f_Q)$ (Fig. 5D).  Thus, near folding occurs at near equilibrium  for $\lambda=\tau_Q/\tau_{hop}\gg\ 1$, while non-equilibrium folding occurs in the opposite limit. Variations in the experimental protocol have to be taken into account in elucidating the pathways explored during the folding process.
\\
\begin{table}
\begin{center}
\caption{\label{tab1}Rate of force-quench in various experiments
}
\begin{tabular}{l||c|c|c|c}
\hline\hline
Experiments &$v_L$ (slew rate) & $\tau_{Q}$ (quench time) & $\tau_F(f_Q)$ (folding time) &\footnote{For $\lambda\ll 1$ we expect non-equilibrium effect which leads to population of long-lived FIMIs. In the opposite limit folding occurs close to equilibrium conditions. 
The values of $\lambda$ for Ub explain the differences in the refolding found in references \cite{FernandezSCI04} and \cite{Schlierf09AngChem}. The value of $\lambda$ in all our simulations are extremely small and cannot be realized in AFM experiments. However, we expect qualitative changes in $R(t)$ only when $\lambda > 1$, and hence our simulation results are qualitatively similar to the small $\lambda (\ll1)$ used in \cite{FernandezSCI04}.}$\lambda=\tau_{Q}/\tau_F(f_Q)$\\ \hline
Poly-ubiquitin \cite{FernandezSCI04} & $\sim 10^4$ $nm/s$& 3 $msec$ & 0.2 $s$ ($f_Q=15$ pN) & 0.015\\
TAR RNA \cite{TinocoBJ06}  & $>200$ $nm/s$& 0.1 $sec$ & $\sim 1$ $sec$&0.1 \\
RNAase H ($I\rightarrow N$) \cite{MarquseeScience05} & $(10-10^3)$ $nm/s$ & ($0.05-5$) $sec$ & 5 $sec$ ($I\rightarrow U$) & $(0.01-1)$\\
RNAase H ($N\rightarrow U$) \cite{MarquseeScience05} & $(10-10^3)$ $nm/s$ & ($0.05-5$) $sec$ &3000 $sec$ ($N\rightarrow U$) & ($1/60000-1/600$) \\
Ubiquitin dimer \cite{Schlierf09AngChem} & 5 $nm/s$ & $\sim$0.16 $sec$ & $\gtrsim$5 $msec$ ($f_Q=0$ pN) & $\lesssim$32\\ 
P5GA hairpin (simulation)& $\sim10^{11}$ $nm/s$& 47 $ps$ & $(10-1000)$ $\mu s$ &$(10^{-9}-10^{-6})$ \\ 
WLC (simulation) & $\sim 10^{13}$ $nm/s$ & 0.5 $p s$ & $\gtrsim 100 \mu s$ & $\lesssim 5\times 10^{-9}$\\ 
\hline
\hline
\end{tabular}
\end{center}
\end{table}

\noindent {\bf METHODS}\\

The P5GA hairpin, as represented using the SOP model (see ref \cite{HyeonBJ07} for the energy function and simulation method), is a two state folder with a transition mid-force of $f_m=14.7$ pN \cite{Hyeon08PNAS}. 
We probe the refolding kinetics for multiple pairs of initial and quench forces, ($f_S$,$f_Q$), whose magnitudes range from $14\leq f_S\leq 70$ pN and $2\leq f_Q\leq 12$ pN (see Fig.1A). The force-quench refolding kinetics for 100 molecules were simulated beginning with an initial tension ($f_S$) and quenched to a final tension ($f_Q$).
The refolding time of the $i^{th}$molecule, $\tau_F(i)$, was defined as the first time the hairpin gained more than 95\% of native contacts after the tension was reduced to $f=f_Q$.
\\

\noindent {\bf SUPPORTING INFORMATION}\\

{\bf $T$-quench versus $f$-quench dynamics : }
Our work also provides insights into 
the qualitative differences between the two-state dynamics upon force or temperature variations (compare $P_{TDE}(R)$ and $P_S(R)$ in Fig.2).  
For an apparent two-state folder, the fraction of folded states is $\varphi_{NBA}(f)=1/[1+e^{-(\Delta F^o_{UF}-f \Delta x_{UF})/k_BT}]$.   
The half-width of fluctuations ($\delta f_{1/2}$) around transition mid-force $f_m=\Delta F_{UF}/\Delta x_{UF}$ using $d\varphi_{NBA}(f)/df$ is  
\begin{equation}
\delta f_{1/2}=\frac{k_BT}{\Delta x_{UF}}\log{\sqrt{\frac{3+2\sqrt{2}}{3-2\sqrt{2}}}}\approx 1.7\times\frac{k_BT}{\Delta x_{UF}}. 
\end{equation}
$\Delta x_{UF}\approx 6$ nm for P5GA hairpin \cite{HyeonBJ07} (see Fig.S3), thus $\delta f_{1/2}\approx 1.2$ pN at $k_BT=4.14$ $pN\cdot nm$. 
For the folded state to be predominantly populated after the force is quenched, the depth of quench should satisfy $\delta f_Q=f_m-f_Q>\delta f_{1/2}\approx 1.2$ pN.
It is noteworthy that only the characteristic force $k_BT/\Delta x_{UF}$ determines the sharpness of the transition ($\delta f_{1/2}$), while $\Delta F_{UF}$ plays no significant role. 
Since $\Delta x_{UF}\sim N$, it follows that $\delta f_{1/2}\sim 1/N$, which differs from the thermodynamic scaling upon temperature variation, $(T_{\Theta}-T)/T_{\Theta} = \delta T/T_{\Theta}\sim 1/\sqrt{N}$ \cite{Brochard81ARPC,deGennes85JPL}. 
Thus, for a given length $N$, the force-induced transition is sharper than the temperature-induced transition \cite{HyeonPNAS05}.  Since $\exp{(f_Q\Delta x^{\ddagger}_{U\rightarrow N}/k_BT)}=\exp{(f_m\Delta x^{\ddagger}_{U\rightarrow N}/k_BT)}\exp{(-\delta f_Q\Delta x^{\ddagger}_{U\rightarrow N}/k_BT)}$, a larger depth of quench exponentially increase the rate of folding by lowering the free energy barrier associated with refolding dynamics from the UBA, in contrast to the homopolymer following a temperature quench ($\tau_c\approx \tau_R \delta T/T_{\Theta}$) \cite{deGennes85JPL}. \\

{\bf Calculation of $P_{RSE}$ and $P_{TSE}$:}
To calculate $P_{RSE}(R)$ over the time traces, we collect the data points from $50<t<70$ $\mu $s for $f_S=56$ pN, and $10<t<30$ $\mu $s for $f_S=14$ pN. Similarily, to calculate $P_{TSE}(R)$ over the time traces, the data are collected from $\tau_F(i)-270<t<\tau_F(i)-250$ $\mu $s for all $f_Q$ values, where $\tau_F(i)$ denotes the folding time of $i^{th}$ trajectory.
\\

{\bf Dependence of folding time of P5GA hairpins on stretch- ($f_S$) and quench-forces ($f_Q$) : }
For a given survival probability ($\Sigma(t)=R(t;f_S,f_Q)/R(0;f_S)$) the average folding time is obtained by using
\begin{equation}
    \tau_F(f_S,f_Q)=\int^{\infty}_0dt\Sigma(t;f_S,f_Q).
  \end{equation}
The force $f_S$ determines the initial distribution of molecules in the UBA and NBA, and is giving the Boltzmann distribution. To probe folding we are interested in the dynamics of molecules that are initially in the UBA. 
  The fraction of molecules in the UBA for the two state RNA hairpin is 
  \begin{eqnarray}
      \varphi_{UBA}(f_S)&=&\frac{1}{1+e^{(\Delta F_{UN}-f_S\Delta x_{UN})/k_BT}}\nonumber\\
        &=&\frac{1}{1+e^{(-\delta f_S\Delta x_{UN})/k_BT}}
      \end{eqnarray}
      where $f_S=f_m+(f_S-f_m)=f_m+\delta f_S$, and $\Delta F_{UN}-f_m\Delta x_{UN}=0$. 
      It follows from Fig.1B in the text that $\tau_f\sim e^{f_Q\Delta x^{\ddagger}_{U\rightarrow N}/k_BT}$.
       Since $f_S$ and $f_Q$ are independent control variables in an experiment, one can factorize the survival probability as $\Sigma(t,f_S,f_Q)=\varphi_{UBA}(f_S)\Sigma(t;\infty,f_Q)$. 
      Therefore, 
      \begin{equation}
	  \tau_F(\delta f_S,\delta f_Q)=\frac{\tau(0)}{1+e^{(-\delta f_S\Delta x_{UN})/k_BT}}e^{-\delta f_Q\Delta x^{\ddagger}_{U\rightarrow N}/k_BT}
	\end{equation}
where $\delta f_S=f_S-f_m$ and $\delta f_Q=f_m-f_Q$.\\

	{\bf The expanding sausage model for a Gaussian chain under tension:  }
	In the presence of an external tension $f$, the simplest modification to the free energy of the expanding sausage model as a function of the length of the sausage ($\approx R$) is given by
	\begin{equation}
	  F=\gamma A-f R\approx \frac{k_BT}{\xi^2}\times 2\pi \rho R-fR=\frac{k_BT}{\xi^2}2\sqrt{\pi\Omega}R^{1/2}-fR,\label{SausageFEq}
	\end{equation}
	where the surface tension for the thermal blob (of size $\xi$) is given as
	$k_BT/\xi^2$, and the exposed surface area of the cylindrical sausage is $A\sim 2\pi \rho R$ ($\rho$ is the radius of the sausage).  The model is schematically shown in Fig. S4A.
	The rate of free energy reduction in the system is given by
	\begin{equation}
	  \dot{F}=\gamma \dot A-f\dot{R}=\left(\frac{k_BT}{\xi^2}\sqrt{\pi\Omega}R^{-1/2}-f\right)\dot{R}, 
	\end{equation}
	while the entropy production due to the dissipation is
	\begin{equation}
	  -T\dot{S}\approx \eta\times R\times \dot{R}^2.
	  \label{eqn:dissipation}
	\end{equation}
	Equating $\dot F=-T\dot S$, as in the main text, we obtain 
	\begin{align}
	  t/\tau_0=\int^{R/R_0}_1\frac{x^{3/2}dx}{\overline{f}^0 x^{1/2}-1}=\frac{2}{(\overline{f}^0)^5}\log{\left(\frac{1-\overline{f}^0\left(\frac{R}{R_0}\right)^{1/2}}{1-\overline{f}^0}\right)}-\sum_{k=1}^4\frac{2}{k(\overline{f}^0)^{5-k}}\left[1-\left(\frac{R}{R_0}\right)^{k/2}\right],
	  \label{eqn:tf}
	\end{align}
	where $R_0$ the initial extension of the polymer, $\tau_0\equiv \frac{\eta\xi^2R_0^{5/2}}{k_BT\sqrt{\pi\Omega}}$, and $\overline{f}^0\equiv f\tau_0/\eta R_0^2$.
	The ratio $R/R_0$ in Eq.\ref{eqn:tf} is interpreted as the survival probability \cite{ZwanzigBook}, $P_S(t)$, as in the main text.
	For $\overline{f}^0\rightarrow 0$, we recover the de Gennes' collapse kinetics in poor solvent without tension,
	$P_S(t)=\left[1-\frac{5}{2}\left(\frac{t}{\tau_0}\right)\right]^{2/5}$.
	The average folding (collapse) time in the absence of tension, given in Eq. 6 of the main text, is
	$\langle\tau\rangle_0=\int^{2/5\times\tau_0}_0dtS_0(t)=\frac{2}{7}\tau_0$.
	Using the definition of $\tau_0$ and $\overline{f}^0$, we can rewrite the free energy $F$ in the dimensionless form,
	\begin{equation}
	  \overline{F}\left(R/R_0\right)=\frac{F(R)\tau_0}{\eta R_0^3}=\left[2(R/R_0)^{1/2}-\overline{f}^0(R/R_0)\right],
	  \label{eqn:freeE}
	\end{equation}
	which is shown in Fig. S4A.

	Above $\overline{f}^0=1$,
	the free energy develops a barrier ($\delta\overline{F}^{\ddagger}$) for refolding, whose position varies with force as $R^{TS}/R_0=1/(\overline{f}^0)^2$.  If $R^{TS}<R_0$, a valid solution of SI Eq. \ref{eqn:tf} cannot be found for all $t>0$ (see also the discussion in the main text).
	The refolding time $\tau_F$ scales with the effective height of a free energy barrier as $\tau_F\sim \exp{(\delta F^{\ddagger})}$ where $\delta F^{\ddagger}$ is easily calculated as $\delta\overline{F}^{\ddagger}=\overline{F}(R^{TS}/R_0)-\overline{F}(1)=1/\overline{f}^0-2+\overline{f}^0$.
	Therefore, the refolding (or nucleation) time scales with the quench force ($\overline{f}^0)$ as
	\begin{equation}
	  \tau_F\sim \exp{\left[\frac{1}{\overline{f}^0}-2+\overline{f}^0\right]},
	\end{equation}
	in parallel to the calculation in the main text.  The survival probability for varying $\overline f^0$ is shown in Fig. S4B.
	The average folding time is determined from Eq. 6, and we see in Fig. S4C that $\tau_F$ increases almost exponentially with the quench force for $\overline f^0\lesssim 0.6$, beyond which there is a sharp increase in $\log[\tau_F(\overline f^0)]$.  However, we note that the rapid reduction in $R$ in the first stage of collapse seen in the simulations (Fig. 1D of the main text) is not observed in the predicted behavior of $R(t)/R_0$ for the Gaussian chain (Fig. S4B).
\\

{\bf Barrier height in the WLC-expanding sausage model for $f_Q>(f_Q)^*$ : }
When $\overline{f}_Q>(\overline{f}_Q)^*$, $R_0/L\rightarrow 1$ and $R^{TS}/L\rightarrow 0$. The conditions of $\overline{F}'(R_0/L)=0$ and $\overline{F}'(R^{TS}/L)=0$
	\begin{align}
	  \overline{F}'(R_0/L)=0&=\left(\frac{R_0}{L}\right)^{-1/2}+\overline{f}_p\left[
	  \frac{1}{4(1-R_0/L)^2}-\frac{1}{4}+\frac{R_0}{L}\right]-\overline{f}_Q\nonumber\\
	  &\approx \overline{f}_p\frac{1}{4(1-R_0/L)^2}-\overline{f}_Q
	\end{align}
	and
	\begin{align}
	  \overline{F}'(R^{TS}/L)=0&=\left(\frac{R^{TS}}{L}\right)^{-1/2}+\overline{f}_p\left[
	  \frac{1}{4(1-R^{TS}/L)^2}-\frac{1}{4}+\frac{R^{TS}}{L}\right]-\overline{f}_Q\nonumber\\
	  &\approx \left(\frac{R^{TS}}{L}\right)^{-1/2}-\overline{f}_Q
	\end{align}
	lead to  
	\begin{align}
	  \frac{R^{TS}}{L}\approx \frac{1}{\overline{f}^2_Q}
	\end{align}
	and 
	\begin{align}
	  \frac{R_0}{L}\approx 1-\frac{1}{2}\sqrt{\frac{\overline{f}_p}{\overline{f}_Q}}.
	\end{align}
	The free energy barrier for $\overline{f}_Q(>(\overline{f}_Q)^*)$ can be computed using
	\begin{align}
	  \overline{F}(R^{TS}/L)&\approx \frac{1}{\overline{f}_Q}+\frac{\overline{f}_p}{4}\frac{1}{\overline{f}_Q^4}\left(\frac{3-2/\overline{f}_Q^2}{1-1/\overline{f}_Q^2}\right)\nonumber\\
	  &\approx \frac{1}{\overline{f}_Q}+\frac{1}{4}\frac{\overline{f}_p}{\overline{f}_Q^4}\left(3+\frac{1}{\overline{f}^2_Q}+\frac{1}{\overline{f}_Q^4}+\cdots\right), 
	\end{align}
	and 
	\begin{align}
	  \overline{F}(R_0/L)&\approx 2(1-\sqrt{s}-\frac{s}{2}-\frac{s^{3/2}}{2}-\cdots)+\frac{\overline{f}_p}{4}\left(\frac{1}{2\sqrt{s}}-6\sqrt{s}+8s\right)-\overline{f}_Q(1-2s)\nonumber\\
	  &=\frac{\overline{f}_p}{8\sqrt{s}}+(2-\overline{f}_Q)-\left(2-\overline{f}_Q+\frac{3}{2}\overline{f}_p\right)\sqrt{s}+2\overline{f}_ps+\mathcal{O}(s^{3/2})\cdots 
	\end{align}
	where $s(\equiv \overline{f}_p/\overline{f}_Q)\rightarrow 0$ when $\overline{f}_Q$ takes large value.
	Thus, the free energy barrier is calculated as
	\begin{align}
	  \delta \overline{F}^{\ddagger}&=\overline{F}(R^{TS}/L)-\overline{F}(R_0/L)\nonumber\\
	  &\approx \left(\frac{1-2\overline{f}_p^2}{\overline{f}_Q}-2+\overline{f}_Q\right)+\left(2-\overline{f}_Q+\frac{3}{2}\overline{f}_p\right)\sqrt{\frac{\overline{f}_p}{\overline{f}_Q}}+\mathcal{O}\left[\left(\frac{\overline{f}_p}{\overline{f}_Q}\right)^{3/2}\right]
	\end{align}
	\\

{\bf The collapse of a semiflexible chain in a poor solvent : } 
	The Hamiltonian used in our simulations of a wormlike chain in a poor solvent undergoing collapse to a toroidal structure upon force quench is given by
	\begin{equation}
	  \mathcal{H}=\frac{k_s}{2a^2}\sum_{i=1}^{N-1}(r_{i,i+1}-a)^2+\frac{k_b}{2}\sum_{i=1}^{N-2}(1-\hat{r}_i\cdot\hat{r}_{i+1})+\epsilon_{LJ}\sum_{i,j}\left[\left(\frac{a}{r_{ij}}\right)^{12}-2\left(\frac{a}{r_{ij}}\right)^6\right]-f(z_{N}-z_1)]\label{WLCHam.eq}
	\end{equation}
	with the parameters, $\epsilon_{LJ}=1.5k_B T$, $k_s=2000k_B T$, $N=200$, $a=0.6$ and $k_b=80$ $k_BT$. 
	To integrate the equation of motion, we used Brownian dynamics algorithm.
	\begin{equation}
	  \vec{r}(t+\Delta t)=\vec{r}(t)-D\vec{\nabla}\mathcal{H}(\{\vec{r}\})\Delta t/k_BT+\vec{R}(t).
	\end{equation}
	where $\vec{R}(t)$ is a vector of Gaussian random number satisfying $\langle \vec{R}\rangle=0$ and $\langle\vec{R}_{\alpha}(t)\cdot\vec{R}_{\beta}(t)\rangle=2D\Delta t\delta_{\alpha\beta}$. We use diffusion coefficient $D=1\times 10^{-7} cm^2/s$, thus characteristic time for simulation is $a^2/2D\approx 2$ $ns$. We chose integration time step $\Delta t=0.00025(a^2/2D)\approx 0.5$ $ps$.\\

\noindent {\bf Acknowledgements : } This work was supported in part by the grants from the National Science Foundation (CHE 05-14056) (D.T.) and the National Research Foundation of Korea (NRF) (No. R01-2008-000-10920-0, KRF-C00142 and KRF-C00180) (C.H.). 
\\

\clearpage
\renewcommand{\thefigure}{S1}
\begin{figure}[ht]
  \includegraphics[width=6.00in]{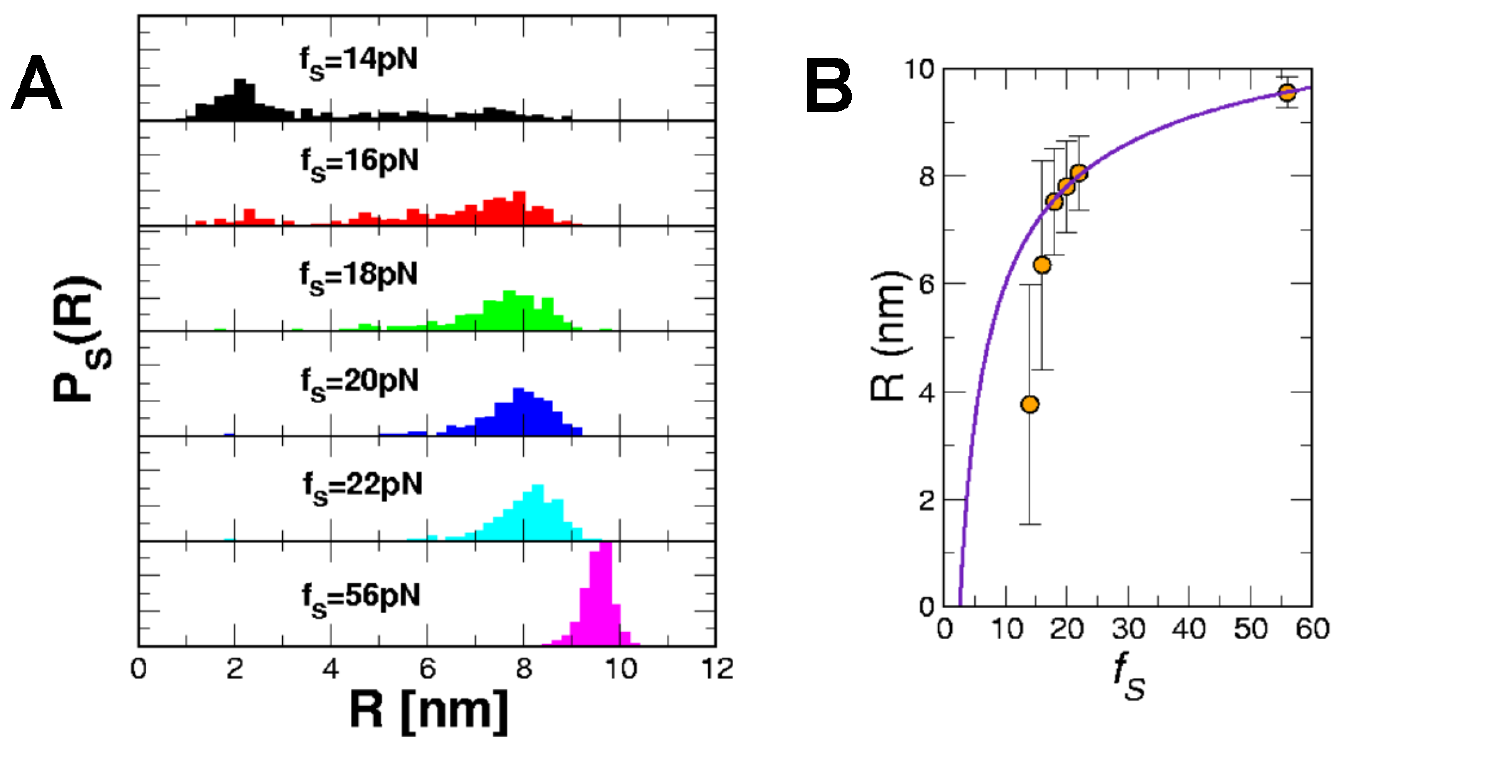}
  \caption{Initial ensemble for the P5GA simulation.
{\bf A} Structural ensembles of the P5GA hairpin represented by
$P(R)$ at each $f_S$.
{\bf B}  The average end-to-end distance as a function $f_S$.
Excluding the values at $f_S=14$ and 16 pN, where the distribution is biomodal, $\langle R\rangle$ vs $f_S$ is well fit by the wormlike chain model
$\langle R\rangle \approx L\left(1-\sqrt{\frac{k_BT}{4l_pf_S}}\right)$, yielding
$L=12.2$ nm and $l_p=0.43$ nm.
\label{FigS1.fig}}
\end{figure}
\renewcommand{\thefigure}{S2}
\begin{figure}[ht]
 \includegraphics[width=6.00in]{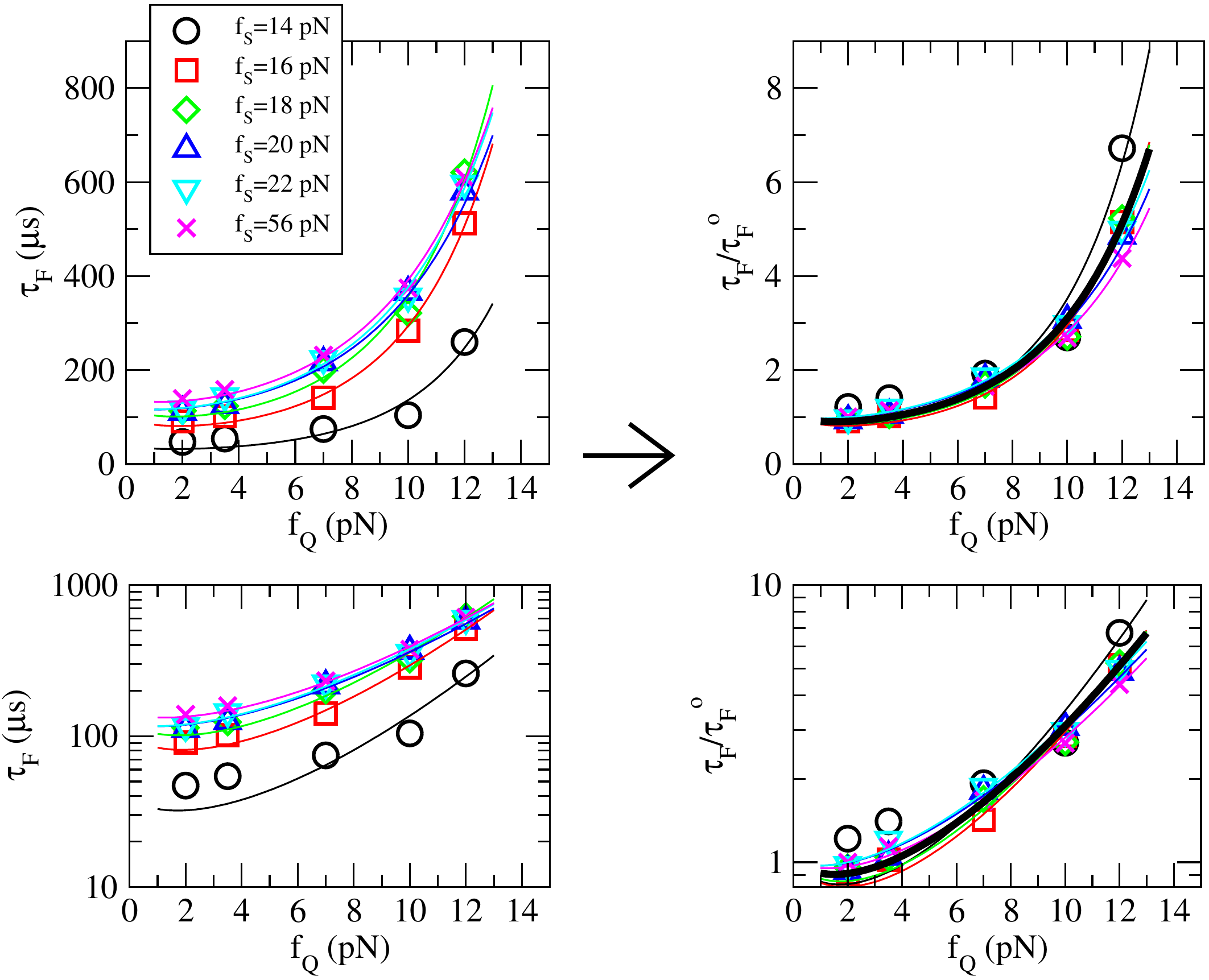}
  \caption{Analysis of $\tau_F$ versus $f_Q$ data by adapting Dudko and coworkers' microscopic model for force spectroscopy 
$
\tau_F(f_Q,f_S)=\tau_F^o(f_S)\left(1+\frac{\nu f_Q\Delta x^{\ddagger}_{U\rightarrow F}}{\Delta F^{\ddagger}_{U\rightarrow F}}\right)^{1-1/\nu}e^{-\Delta F^{\ddagger}_{U\rightarrow F}/k_BT\cdot[1-(1+\nu f_Q\Delta x^{\ddagger}_{U\rightarrow F}/\Delta F^{\ddagger}_{U\rightarrow F})^{1/\nu}]}
$ 
where we use $\nu=2/3$ (cubic potential) and change the sign of force from the one in Ref.\cite{Dudko08PNAS} to consider the refolding process under tension.   
When $\tau_F$ is scaled by $\tau_F^o$ for each $f_S$, the folding times approximately collapse to a single curve (right). 
From the fit using thick line, the parameters extracted for the collapsed data are $\Delta x^{\ddagger}=0.45$ nm\ and $\Delta F^{\ddagger}_{U\rightarrow F}=0.56$ $pN\cdot nm$. 
\label{FigS2.fig}}
\end{figure}
\renewcommand{\thefigure}{S3}
\begin{figure}[ht]
  \includegraphics[width=4.00in]{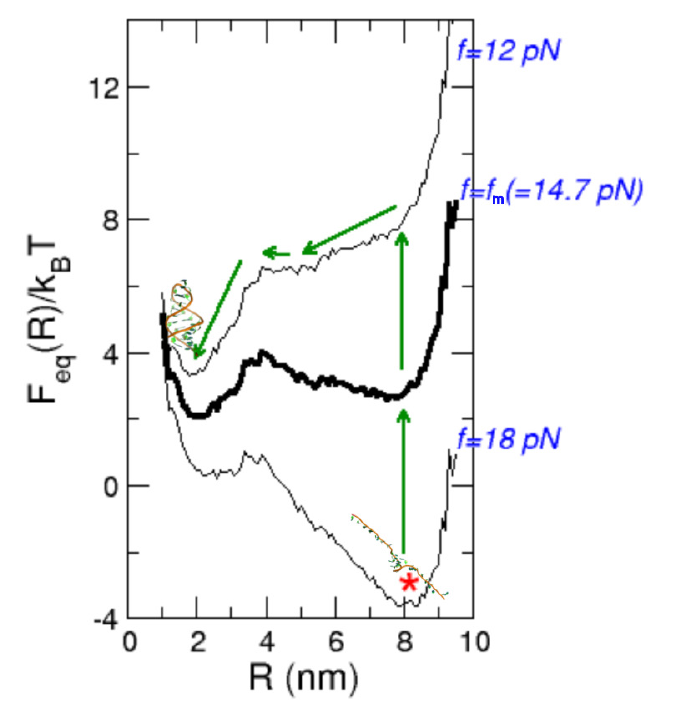}
  \caption{The free energy of the P5GA hairpin as a function of $R$ near $f=f_m$, schematically illustrating how the initially stretched structure of RNA at $f_S=18$ pN adapts its structure under the force-quench condition at $f_Q=12$ pN.  \label{FigS3.fig}}
\end{figure}
\renewcommand{\thefigure}{S4}
\begin{figure}[ht]
  \includegraphics[width=6.00in]{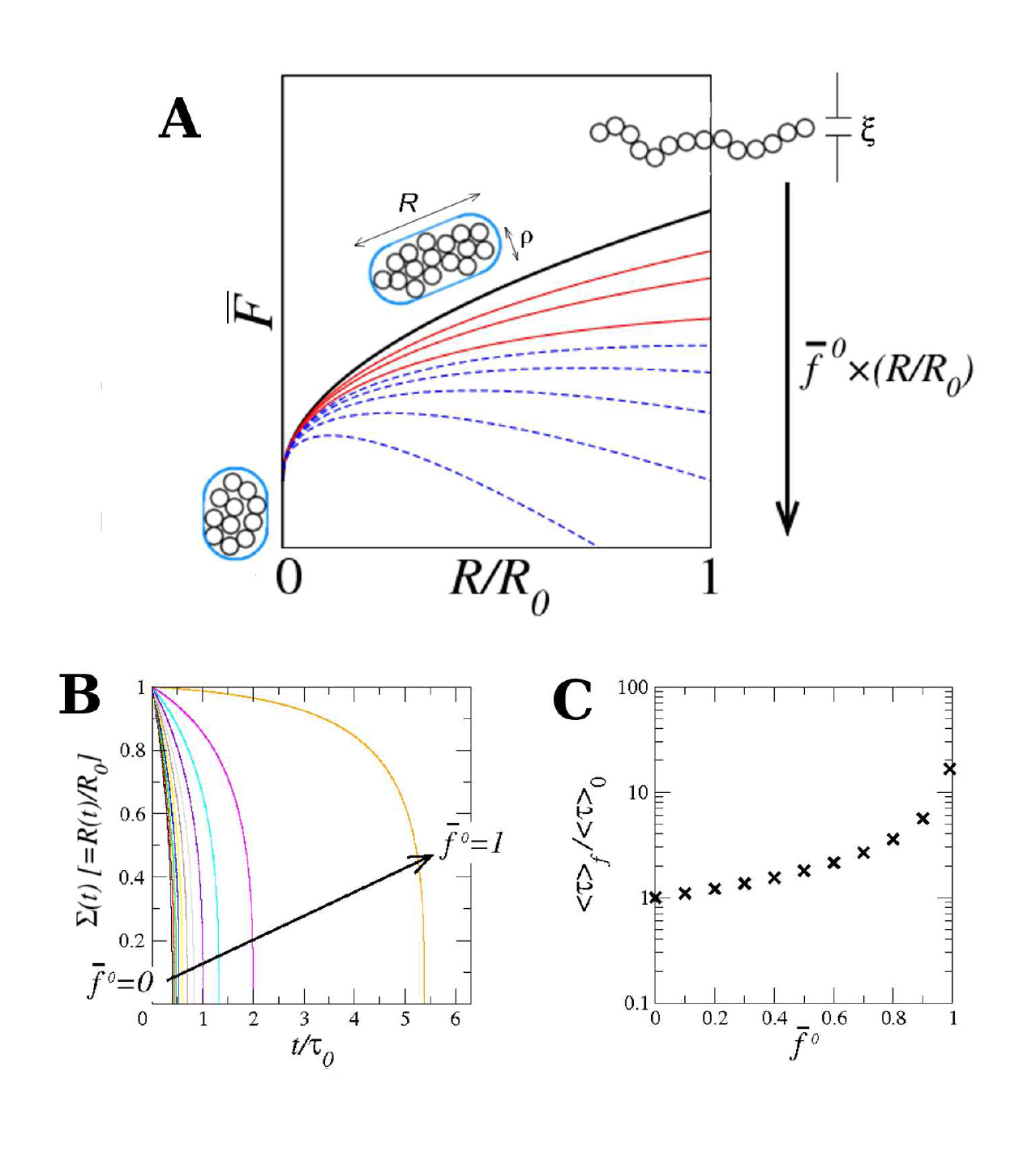}
  \caption{The expanding sausage model under variation of the external tension. {\bf A}. The free energy profile calculated using $R/R_0$ with varying tension ($f_Q$). The collapse of a homopolymer is diagrammed schematically for $f_Q=0$, but the extended structure can be stabilized in the presence of tension.
{\bf B}. Reduction of the molecular extension with increasing tension (from left to right).
{\bf C}. The average refolding time as a function of external tension, with $\log(\tau_F)\sim\overline{f}^0$ for $\overline{f}^0\lesssim 0.6$.
\label{FigS4.fig}}
\end{figure}
\renewcommand{\thefigure}{S5}
\begin{figure}[ht]
  \includegraphics[width=6.00in]{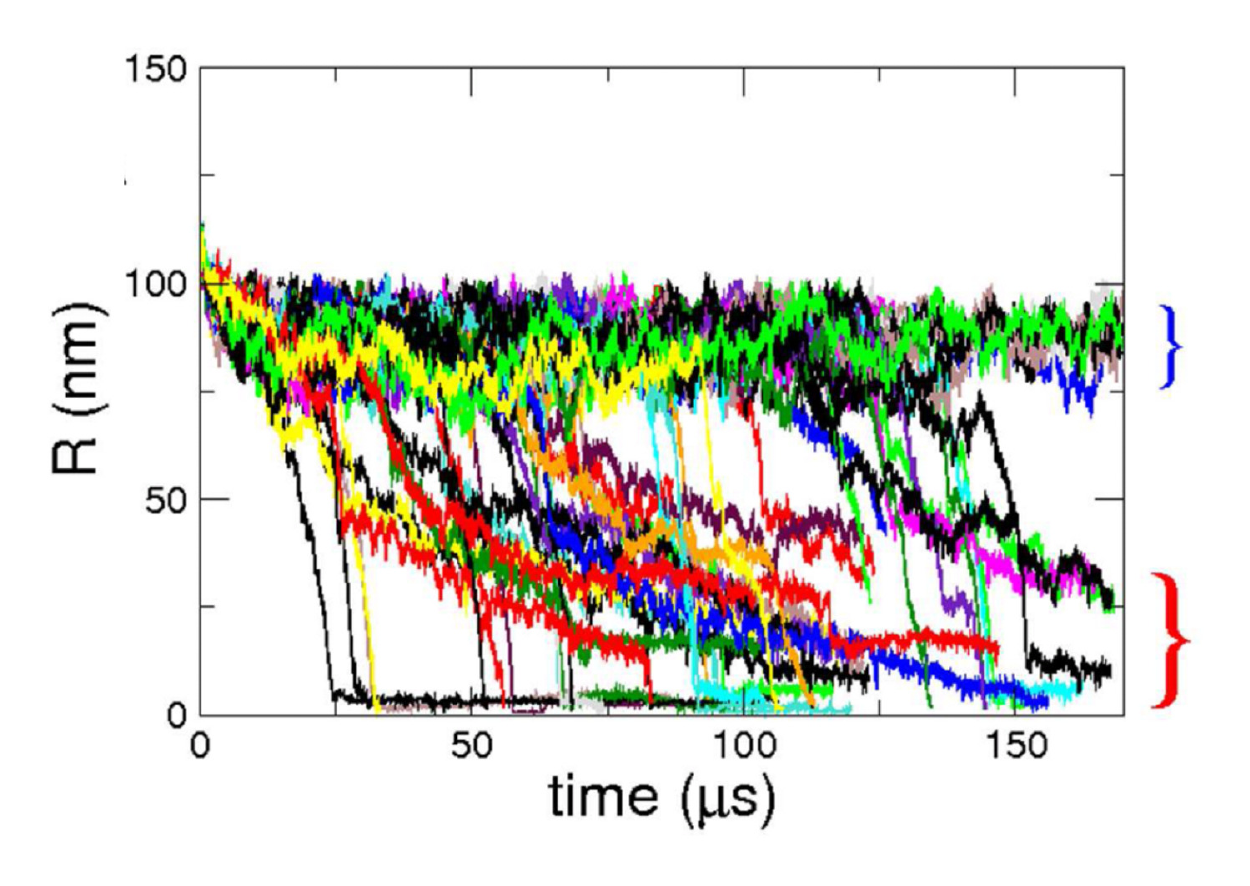}
  \caption{Full 50 force-quench induced dynamic trajectories of semiflexible polymers. 
\label{FigS5.fig}}
\end{figure}
\renewcommand{\thefigure}{S6}
\begin{figure}[ht] 
  \includegraphics[width=6.00in]{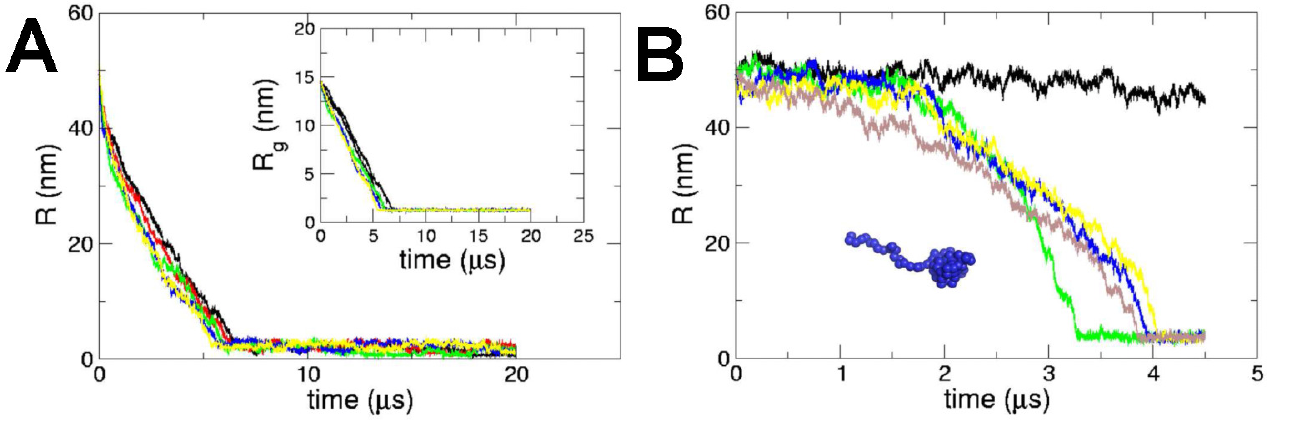}
  \caption{The collapse dynamics of a freely jointed homopolymer in a poor solvent is shown, with $f_S=83$ pN to $f_Q=4$ pN ({\bf A}) and $f_S=83$ pN to $f_Q=75$ pN ({\bf B}). The flexible chain shows no evidence of the plateau for small $f_Q$, but a higher quench force stabilizes intermediate structures.\label{FigS6.fig}}
\end{figure}


\end{document}